\documentclass[default,iicol]{sn-jnl-corrected}

\jyear{2023}%

\usepackage{epstopdf}
\usepackage{color}
\usepackage{ulem}
\usepackage{amsmath}
\usepackage{amssymb}
\usepackage{bm}
\newcommand{\abs}[1]{\vert #1\vert}

\renewcommand{\vec}[1]{\bm{#1}}

\newcommand{\pv}{\vec{p}}

\newcommand{\rv}{\vec{r}}

\newcommand{\Id}{{\mathsf{I}}}
\newcommand{\calR}{\mathcal{R}}
\newcommand{\calH}{\mathcal{H}}
\newcommand{\calF}{\mathcal{F}}
\newcommand{\calI}{\mathcal{I}}

\begin{document}
\title[Corrections to LDA for superfluid trapped fermions from the Wigner-Kirkwood $\hbar$ expansion]{Corrections to Local Density Approximation for
  superfluid trapped fermionic atoms from the Wigner-Kirkwood $\hbar$ expansion}
\author[1,2]{\fnm{Peter} \sur{Schuck}}
\author*[1]{\fnm{Michael} \sur{Urban}} \email{michael.urban@ijclab.in2p3.fr}
\author*[3,4]{\fnm{Xavier} \sur{Vi\~nas}} \email{xavier@fqa.ub.edu}
\affil*[1]{\orgname{Universit\'e Paris-Saclay, CNRS/IN2P3, IJCLab},
  \orgaddress{\city{Orsay}, \postcode{F-91405}, \country{France}}}
\affil[2]{\orgname{Universit\'e Grenoble Alpes, CNRS, LPMMC},
  \orgaddress{\city{Grenoble}, \postcode{F-38000}, \country{France}}}
\affil*[3]{\orgdiv{Departament de F\'isica Qu\`antica i Astrof\'isica and
    Institut de Ci\`encies del Cosmos},
  \orgname{Facultat de F\'isica, Universitat de Barcelona},
  \orgaddress{\street{Diagonal 645}, \city{Barcelona}, \postcode{E-08028},
    \country{Spain}}}
\affil[4]{\orgname{Institut Menorqu\'{\i} d'Estudis},
  \orgaddress{\street{Cam\'{\i} des Castell 28},
    \city{Ma\'o}, \postcode{E-07702}, \country{Spain}}}


\abstract{A semiclassical second-order differential equation for the 
inhomogeneous local gap $\Delta(\rv)$ is derived from a strict second-order 
$\hbar$ expansion of the anomalous pairing tensor and compared with a similar 
equation given by Simonucci \textit{et al.} in \cite{simonucci14}. The 
second-order normal density matrix is given as well. Several extra gradient terms
are revealed. Second-order expressions at finite temperature are given for the 
first time. The corresponding Ginzburg-Landau equation is presented and it is 
shown that, compared to the equation of Baranov and Petrov \cite{B-P-98}, an extra
second-order gradient term is present. Applications to the pairing gap in cold
atoms in a harmonic trap are presented.}
%

\maketitle
%
\section{Introduction}
\label{intro}
The solution of Hartree-Fock-Bogoliubov (HFB) \cite{Rin80} or 
Bogoliubov-de Gennes (BdG) \cite{Gen66} equations for finite systems
like superfluid nuclei or cold
atoms in traps still can be a source of a numerical challenge, in
particular if the particle number is very large and in the absence of
spatial symmetries. In these cases the recourse to semiclassical
methods can be of valuable help. In this article we will present such
a formalism based essentially on a Thomas-Fermi like approach
generalized to the superfluid case. This will be achieved with the
Wigner-Kirkwood $\hbar$ expansion of the various density matrices.

The Wigner-Kirkwood $\hbar$ expansion of the single-particle density
matrix in the normal-fluid situation is well known and has been
applied many times, e.g., to finite nuclei \cite{Rin80}. The $\hbar$
expansion of the generalized density matrix in the superfluid case has
been considered only in very few works and has practically not found
any applications so far.

The lowest order of the Wigner-Kirkwood expansion in the superfluid
case corresponds to the well known Local-Density Approximation (LDA),
which treats the system in each point $\rv$ as if it was uniform
matter of density $\rho(\rv)$. In the special case of cold atoms in
the weak-coupling regime and at zero temperature, this gives, see,
e.g. \cite{grasso03}
\begin{equation}
  \Delta(\rv)
  = 8 \mu(\rv) \exp \Big(-2 - \frac{\pi}{2 k_F(\rv)\abs{a_F}}\Big),
\label{gaplda}
\end{equation}
where $\mu(\rv)$ and $k_F(\rv)$ are the position dependent chemical
potential and Fermi momentum, respectively, and $a_F$ is the
scattering length.

The full $\hbar^2$ correction to the generalized density matrix of the
HFB equations has been derived for the first time almost 30 years ago
by Taruishi and Schuck \cite{taruishi92}. Since then, it has been
reconsidered by Ullrich and Gross \cite{Gross}, Csordas \textit{et
  al.} \cite{csordas10}, and more recently by Pei \textit{et al.}
\cite{pei15}. The found expressions are relatively complex with many
gradient terms of different kinds. Independently of these works,
Simonucci and Strinati recently derived a relatively simple
second-order differential equation for the local gap not from an
$\hbar$ expansion technique but applying some coarse graining method
to the BdG equations \cite{simonucci14}. They dubbed their method
`Local Phase Density Approximation (LPDA)' and the corresponding
differential equation reads as
\begin{equation}
- \frac{m}{4 \pi \hbar^2 a_F} \Delta(\rv) = \calI_0(\rv) \Delta(\rv)
+ \calI_1(\rv)\frac{\hbar^2}{4m} \nabla^2 \Delta(\rv),
\label{hfb1}
\end{equation}
where
$m$ is the fermion mass, and $\calI_0$ and $\calI_1$ are some
functions of $\rv$, $\Delta$, and temperature $T$ to be given below in
the main text. This equation has been applied with great success to
vortex creation in rotating cold atom traps \cite{Sim15}. Even earlier
Baranov and Petrov derived a Ginzburg-Landau (GL) equation applicable
to cold atoms in harmonic traps \cite{B-P-98} also using gradient
expansion techniques. In this paper we will show how those equations
are related to a strict $\hbar$ expansion of the Wigner-Kirkwood type
of the generalized density matrix for inhomogeneous superfluid
systems.

The paper is organized as follows. In the next section, we will
summarize the $\hbar$ expansion to second order. In section
\ref{sec:gl} we derive the corresponding GL equation that is valid
close to the critical temperature $T_c$. In section \ref{sec:results}
we numerically implement the different approximations for the case of
fermionic atoms in a harmonic trap and compare the semiclassical
results to full HFB calculations.  Conclusions and further discussions
are given in section \ref{sec:conclusions}. An extended Appendix is
presented at the end.

Let us mention that the present paper was one of the last works of
Peter Schuck and unfortunately he was not able to see the final
version.
\section{The full $\hbar$ expansion}
\label{sec:hbar2}
\subsection{Second order $\hbar$ expressions of pairing tensor and
  normal density matrix}
In LDA or Thomas-Fermi approximation, i.e., to zeroth order in
$\hbar$, the Wigner transforms of the pairing tensor and of the normal
density matrix are simply given by their well known respective
expressions in uniform matter,
\begin{gather}
  \kappa_0(\rv,\pv) = \frac{\Delta}{2E}(1-2f(E))\,,
  \label{pairdent0}\\
  \rho_0(\rv,\pv) = \frac{1}{2}\Big[1 - \frac{h}{E}(1-2f(E))\Big]\,,
  \label{normdent0}
\end{gather}
with the only difference that the single-particle hamiltonian
$h(\rv,\pv) = p^2/(2m^*(\rv)) + U(\rv) - \mu$, gap $\Delta(\rv)$, and
quasiparticle energy $E(\rv,\pv) = \sqrt{h(\rv,\pv)^2+\Delta(\rv)^2}$
now depend not only on the momentum $\pv$ but also on the spatial
coordinate $\rv$. Note that the single-particle potential $U(\rv)$
may include some mean-field potential in addition to the external
(trap) potential $V(\rv)$. From now on, we define $\mu(\rv) =
\mu-U(\rv)$, and we also allow for the possibility of a density
dependent effective mass $m^*(\rv) = m/\gamma(\rv)$. The function
$f(E) = (e^{E/T}+1)^{-1}$ in eqs. (\ref{pairdent0}) and
(\ref{normdent0}) is the Fermi function.

The second-order correction to the pairing density in phase space for
inhomogeneous Fermi systems was first derived by Taruishi and Schuck
in \cite{taruishi92} from the Wigner-Kirkwood $\hbar$ expansion of the
HFB Bloch propagator \cite{taruishi92}. Later on, this pairing density
has also been obtained from the $\hbar$ expansion of the Green's
function of the HFB equation \cite{pei15} in the $T=0$ limit. A third
derivation with some applications is given in \cite{csordas10}. A
further derivation having the merit to consider the pairing tensor as
a complex quantity, necessary when one is to include a vector
potential as a magnetic field or rotation, can be found in
\cite{Gross}. For completeness, the generalized density matrix of
superfluid systems is given in appendix \ref{app:R}.

The pairing density in phase space with gradient corrections at finite
temperature and without vector potential can be written as
$\kappa({\rv},{\pv}) = \kappa_0({\rv},{\pv}) + \kappa_2({\rv},{\pv})$
\cite{taruishi92,Gross}, where the $\hbar^2$ contribution reads
\begin{equation}
  \kappa_2(\rv,\pv) = \sum_{i=1,2,4,5,6,7,10}
    c^\kappa_i(\rv,\pv) \frac{\hbar^2}{m} f_i(\rv,\pv)\,,
    \label{pairdent2a}
\end{equation}
with
\begin{align}
c^\kappa_1 =&
  -\tfrac{\Delta(2h^2-\Delta^2)}{32E^5}(1-2f)
  - \tfrac{\Delta(2h^2-\Delta^2)}{16E^4}f'\nonumber\\
 &+ \tfrac{\Delta h^2}{16E^3}f''\,,\nonumber\\
c^\kappa_2 =&
  \tfrac{h(h^2-2\Delta^2)}{16E^5}(1-2f)
  + \tfrac{h(h^2-2\Delta^2)}{8E^4}f'
  + \tfrac{h\Delta^2}{8E^3}f''\,,\nonumber\\
c^\kappa_4 =&
  \tfrac{h\Delta(2h^2-3\Delta^2)}{16E^7}(1-2f)
  + \tfrac{h\Delta(2h^2-3\Delta^2)}{8E^6}f'\nonumber\\
 &- \tfrac{h\Delta(h^2-\Delta^2)}{8E^5}f''
  + \tfrac{h^3\Delta}{24E^4}f'''\,,\nonumber\\
c^\kappa_5 =&
  -\tfrac{h^4+\Delta^4-3h^2\Delta^2}{8E^7}(1-2f)
  - \tfrac{h^4+\Delta^4-3h^2\Delta^2}{4E^6}f'\nonumber\\
 &- \tfrac{h^2\Delta^2}{2E^5}f''
  + \tfrac{h^2\Delta^2}{12E^4}f'''\,,\nonumber\\
c^\kappa_6 =&
  -\tfrac{h\Delta(3h^2-2\Delta^2)}{16E^7}(1-2f)
  - \tfrac{h\Delta(3h^2-2\Delta^2)}{8E^6}f'\nonumber\\
 &+ \tfrac{h\Delta(h^2-\Delta^2)}{8E^5}f''
  + \tfrac{h\Delta^3}{24E^4}f'''\,,\nonumber\\
c^\kappa_7 =&
  \tfrac{5h^2\Delta^2}{16E^7}(1-2f)
  + \tfrac{5h^2\Delta^2}{8E^6}f'
  + \tfrac{h^4+\Delta^4}{8E^5}f''\nonumber\\
 &+ \tfrac{h^2\Delta^2}{24E^4}f'''\,,\nonumber\\
c^\kappa_{10} =&
  -\tfrac{\Delta}{16E^5}(1-2f)
  - \tfrac{\Delta}{8E^4}f'
  + \tfrac{\Delta}{24E^2}f'''\,.
\label{pairdent2}
\end{align}
Here, we have used the notation $f = f(E), f' = \partial f(E)/\partial
E$, etc. Since, for $\Delta > 0$, one has necessarily $E>0$, the
zero-temperature limit in the superfluid phase is easily obtained by
setting $f = f'= f''= f'''=0.$ The functions $f_i(\rv,\pv)$ are
combinations of spatial gradients up to second order of the potential
$U(\rv)$, the inverse effective mass $\gamma(\rv)=m/m^*(\rv)$, and the
gap $\Delta(\rv)$. After averaging over the direction of $\pv$, they
are given by
\begin{align}
f_1 =& \frac{\gamma p^2}{m}\nabla^2 \gamma
  + 2\gamma \nabla^2 U - \frac{2}{3} \frac{p^2}{m} (\nabla \gamma)^2
  \,,\nonumber\\
f_2 =& \gamma \nabla^2 \Delta
  \,,\nonumber\\
f_4 =& -\frac{1}{12}\frac{\gamma p^4}{m^2}(\nabla \gamma)^2
  + \gamma (\nabla U)^2
  + \frac{1}{6}\frac{\gamma^2 p^4}{m^2} \nabla^2 \gamma
  \,,\nonumber\\
  &+ \frac{1}{3}\frac{\gamma^2 p^2}{m} \nabla^2 U
  + \frac{1}{3}\frac{\gamma p^2}{m} \nabla \gamma \cdot \nabla U
  \,,\nonumber\\
f_5 =& \frac{1}{6}\frac{\gamma p^2}{m} \nabla \gamma \cdot \nabla \Delta
  + \gamma \nabla U \cdot \nabla \Delta \nonumber \\
f_6 =& \gamma (\nabla \Delta)^2 \quad\nonumber \\
f_7 =& \frac{1}{3}\frac{\gamma^2 p^2}{m} \nabla^2 \Delta \nonumber\\
f_{10} =& \frac{1}{3}\frac{\gamma^2 p^2}{m} (\nabla \Delta)^2
\label{functionsfi}
\end{align}
These functions were derived first in \cite{taruishi92} and also given
explicitly in \cite{pei15, csordas10, Gross}.

Analogously one can also derive the finite-temperature expression of
the correction to the normal density matrix \cite{pei15, taruishi92,
  csordas10,Gross}
\begin{equation}
  \rho_2(\rv,\pv) = \sum_{i=1,2,4,5,6,7,10}
    c^\rho_i(\rv,\pv) \frac{\hbar^2}{m} f_i(\rv,\pv)\,,
\end{equation}
with
\begin{align}
c^\rho_1 =&
  -\tfrac{3h\Delta^2}{32E^5}(1-2f)
  -\tfrac{3h\Delta^2}{16E^4}f'
  -\tfrac{h^3}{16E^3}f''\,,\nonumber\\
c^\rho_2 =&
  \tfrac{\Delta(2h^2-\Delta^2)}{16E^5}(1-2f)
  +\tfrac{\Delta(2h^2-\Delta^2)}{8E^4}f'\nonumber\\
 &-\tfrac{h^2\Delta}{8E^3}f''\,,\nonumber\\
c^\rho_4 =&
  \tfrac{\Delta^2(4h^2-\Delta^2)}{16E^7}(1-2f)
  +\tfrac{\Delta^2(4h^2-\Delta^2)}{8E^6}f'\nonumber\\
 &-\tfrac{h^2\Delta^2}{4E^5}f''
  -\tfrac{h^4}{24E^4}f'''\,,\nonumber\\
c^\rho_5 =&
  -\tfrac{h\Delta(2h^2-3\Delta^2)}{8E^7}(1-2f)
  -\tfrac{h\Delta(2h^2-3\Delta^2)}{4E^6}f'\nonumber\\
 &+\tfrac{h\Delta(h^2-\Delta^2)}{4E^5}f''
  -\tfrac{h^3\Delta}{12E^4}f'''\,,\nonumber\\
c^\rho_6 =&
  -\tfrac{5h^2\Delta^2}{16E^7}(1-2f)
  -\tfrac{5h^2\Delta^2}{8E^6}f'
  -\tfrac{h^4+\Delta^4}{8E^5}f''\nonumber\\
 &-\tfrac{h^2\Delta^2}{24E^4}f'''\,,\nonumber\\
c^\rho_7 =&
  -\tfrac{h\Delta(2h^2-3\Delta^2)}{16E^7}(1-2f)
  -\tfrac{h\Delta(2h^2-3\Delta^2)}{8E^6}f'\nonumber\\
 &+\tfrac{h\Delta(h^2-\Delta^2)}{8E^5}f''
  -\tfrac{h^3\Delta}{24E^4}f'''\,,\nonumber\\
c^\rho_{10} =&
  \tfrac{h}{16E^5}(1-2f)
  +\tfrac{h}{8E^4}f'
  -\tfrac{h}{24E^2}f'''\,.
\label{normdent2}
\end{align}

\subsection{Local Densities and the LPDA}
From the density matrices in phase space, the local pairing and normal
densities can be obtained by integrating over momentum, e.g.,
\begin{equation}
  \kappa(\rv) = \int\!\!\frac{d^3p}{(2\pi\hbar)^3}\kappa(\rv,\pv)\,,
\end{equation}
and analogously for $\rho(\rv)$. In the particular case of a contact
pairing force (implicitly assumed because $\Delta$ is taken momentum
independent), the gap equation reads
\begin{equation}
  -\frac{1}{g}\Delta(\rv) = \kappa(\rv)\,,
  \label{gapequation}
\end{equation}
with the coupling constant $g = 4\pi\hbar^2 a_F/m$, if the divergence
of the momentum integral of $\kappa_0(\rv,\pv)$ is regularized in the
standard way \cite{SadeMelo} by replacing $\kappa_0(\rv,\pv)\to
\kappa_0(\rv,\pv)-m\Delta(\rv)/p^2$. In LDA, this
gives
\begin{equation}
  \kappa_0(\rv) = \Delta(\rv)\calI_0(\rv)\,,
\end{equation}
where
\begin{equation}
  \calI_0(\rv) = \int\!\!\frac{d^3p}{(2\pi\hbar)^3}
  \Big(\frac{1-2f}{2E}-\frac{m}{p^2}\Big)
  \label{I0}
\end{equation}
is the integral appearing in eq. (\ref{hfb1}) according to the
definition of \cite{simonucci14}.

Let us now consider the $\hbar^2$ correction to the local pairing
density, $\kappa_2(\rv)$. If we look, e.g., at the terms proportional
to $\nabla^2\Delta(\rv)$, which we shall write as $Y^\kappa_1(\rv)
\nabla^2\Delta(\rv)$, we see from the definitions (\ref{functionsfi})
of the $f_i$ functions that we must take into account in
eq. (\ref{pairdent2a}) the terms of $\kappa_2(\rv,\pv)$ that multiply
$f_2$ and $f_7$, i.e., $c^\kappa_2$ and $c^\kappa_7$, and that the
coefficient $Y^\kappa_1(\rv)$ is given by
\begin{equation}
    Y^\kappa_1(\rv) = \frac{\hbar^2}{m^*}\int\!\!
    \frac{d^3p}{(2\pi\hbar)^3} \Big(c^\kappa_2(\rv,\pv) +
    \frac{p^2}{3m^*}c^\kappa_7(\rv,\pv)\Big)\,.
    \label{Y1hbar2}
\end{equation}
(remember that $\gamma = m/m^*$). Similarly, one can collect the other
combinations of gradients contained in the $f_i$ functions and one
finds that the local pair density $\kappa_2(\rv)$ can be written in
the form
\begin{multline}
  \kappa_2(\rv)=Y^\kappa_1(\rv) \nabla^2 \Delta
  + Y^\kappa_2(\rv)(\nabla \Delta)^2 + Y^\kappa_3(\rv) \nabla^2 U \\
  + Y^\kappa_4(\rv)(\nabla U)^2 + Y^\kappa_7(\rv) \nabla U \cdot\nabla\Delta\\
  + Y^\kappa_5(\rv) \frac{\nabla^2 \gamma}{\gamma}
  + Y^\kappa_6(\rv) \Big(\frac{\nabla \gamma}{\gamma}\Big)^2\\
  + Y^\kappa_8(\rv)\frac{\nabla \gamma \cdot \nabla U}{\gamma}
  + Y^\kappa_9(\rv)\frac{\nabla \gamma \cdot \nabla \Delta}{\gamma}\,,
  \label{pairden1}
\end{multline}
where the $Y^\kappa_i(\rv)$ are obtained by integrating the
corresponding terms of eq. (\ref{pairdent2}) over the momentum
$\pv$. Notice that they depend on $\rv$ only through their dependence
on $\mu(\rv)$ and $\Delta(\rv)$. In the case $m^* = m$ (i.e., $\gamma
= 1$), only the terms in the first two lines of eq. (\ref{pairden1})
contribute. The correction to the normal density, $\rho_2(\rv)$, can
be written analogously, with functions $Y^\rho_i(\rv)$ instead of
$Y^\kappa_i(\rv)$.

At finite temperature, the integrations over momentum for the
functions $Y^\kappa_i$ and $Y^\rho_i$ must be done numerically. But in
the $T \to 0$ limit, the semiclassical pairing and normal densities
can be integrated analytically over the momentum and expressed in
terms of complete elliptic integrals. For completeness, the analytical
expressions for the functions $Y^\kappa_i$ and $Y^\rho_i$ at $T=0$ are
given in appendix \ref{app:yi}.

The LPDA equation (\ref{hfb1}) derived in \cite{simonucci14} is
contained in the first term on the r.h.s. of eq. (\ref{pairden1}) if
we identify $\calI_1 = 4m Y^\kappa_1/\hbar^2$. We see that this coarse
graining method just picks one of several $\hbar$ correction terms of
the full expression of the pairing density (\ref{pairden1}). It can be
guessed that it is the most important term since it is the one where
the Laplacian acts directly on the gap $\Delta(\rv)$ in (\ref{hfb1})
and (\ref{pairden1}). Let us now compare our result (\ref{Y1hbar2})
for the function $Y^\kappa_1$ with the corresponding coefficient in
the LPDA,
\begin{equation}
    Y^\kappa_{1,\text{LPDA}}(\rv) = \frac{\hbar^2}{4m}\calI_1(\rv)\,,
    \label{Y1LPDA}
\end{equation}
with
\begin{equation}
    \calI_1(\rv) = \int\!\! \frac{d^3p}{(2\pi\hbar)^3}
    \bigg[\frac{h(1-2f)}{4E^3}+\frac{hf'}{2E^2} +
      \frac{p^2f''}{6mE}\bigg]\,.
    \label{I1}
\end{equation}
as given after eq. (13) in \cite{simonucci14}. For historical reasons,
we keep the notation $h \equiv h(\rv,\pv)$ instead of $\xi$ used in
\cite{simonucci14}. Inserting the explicit expressions for
$c^\kappa_2$ and $c^\kappa_7$ given in eq. (\ref{pairdent2}) into
eq. (\ref{Y1hbar2}), we see that our expression for $Y^\kappa_1$ is
different from $Y^\kappa_{1,\text{LPDA}}$. But as we will see in the
next section, at least near the critical temperature $T_c$ they become
equal. Notice also that at zero temperature, $Y_1$ diverges in the
limit $\Delta\to 0$, whereas $Y_{1,\text{LPDA}}$ remains finite. Also
other $Y^\kappa_i$ coefficients present divergences for $\Delta\to 0$
at zero temperature. At finite temperature, this problem does not
exist.

\section{Generalized Ginzburg-Landau equation}
\label{sec:gl}
We now want to write the $\hbar^2$ contribution to the pairing density
for temperatures close to the critical one. In this regime we retain
in eqs. (\ref{pairdent2a}) and (\ref{pairdent2}) only linear terms in
$\Delta$. As it can be seen in eq. (\ref{functionsfi}), $f_1$ and
$f_4$ are independent of $\Delta$, $f_2$, $f_5$ and $f_7$ are linear
in $\Delta$, and $f_6$ and $f_{10}$ are quadratic in $\Delta$ and
therefore $c^\kappa_6$ and $c^\kappa_{10}$ do not contribute. In this
limit, after retaining in the remaining contributions only terms
linear in $\Delta$, the surviving terms in eq. (\ref{pairdent2}) read
\begin{align}
  c^\kappa_1 =& - \Delta \bigg[\frac{1-2f}{16E^3}+
    \frac{f'}{8E^2} - \frac{f''}{16E}\bigg]\,,
  \nonumber\\
  c^\kappa_2 =& \frac{h(1-2f)}{16E^3} + \frac{hf'}{8E^2}\,,
  \nonumber\\
  c^\kappa_4 =& \Delta\bigg[\frac{h(1-2f)}{8E^5} + \frac{hf'}{4E^4}
    - \frac{hf''}{8E^3} + \frac{hf'''}{24E^2}\bigg]\,,
  \nonumber\\
  c^\kappa_5 =& -\bigg[\frac{1-2f}{8E^3} + \frac{f'}{4E^2}\bigg]\,,
  \nonumber\\
  c^\kappa_7 =& \frac{f''}{8E}\,,
\label{pairdent3}
\end{align}
to be evaluated in the limit $E\to \abs{h}$.

One may ask the question of the relation of this linearized expression
with the GL equation. In \cite{simonucci14} the equivalence with the
GL equation is demonstrated for the terms proportional to
$\nabla^2\Delta$. The expression (\ref{pairdent2a}) with eq.
(\ref{pairdent3}) contains, however, more gradient terms than the
original GL equation. They result from gradients of the mean-field
potential and effective mass. We want to first reconsider the terms
already treated in \cite{simonucci14}, which contain $\nabla^2\Delta$,
i.e., the terms with $f_2$ and $f_7$.

The corresponding term in the local pair density $\kappa(\rv)$ written
in the form of eq. (\ref{pairden1}) is $Y^\kappa_1(\rv)
\nabla^2\Delta$. Inserting eq. (\ref{pairdent3}) into the expression
(\ref{Y1hbar2}) for $Y^\kappa_1$, we find
\begin{equation}
  Y^\kappa_1 = \frac{\hbar^2}{m^*}\int\!\!\frac{d^3p}{(2\pi\hbar)^3}
    \bigg[\frac{h(1-2f)}{16E^3}+\frac{hf'}{8E^2} +
      \frac{p^2f''}{24m^*E}\bigg]\,.
\end{equation}
Comparing this expression with eq. (\ref{I1}), we see that in the
limit $T\to T_c$, i.e., $\Delta\to 0$, the coefficient $Y^\kappa_1$
obtained within the $\hbar$ expansion coincides with the one of
\cite{simonucci14} obtained within the LPDA, $Y^\kappa_{1,\text{LPDA}}
= \hbar^2/(4m)\calI_1$, if we replace in the latter $m\to m^*$.

We now take the limit $T \to T_c$ and, thus, $\Delta \to 0$, and make
the change of variables to $x = h/(2T)$. Considering the weak-coupling
regime where $-\mu(\rv)/(2T_c)$ is very negative, we can extend
the lower limit of the integral to $-\infty$ and neglect the first two
terms on the r.h.s. of (\ref{I1}) which stem from $c^\kappa_2$
(their integrand is odd in $x$ with $\sqrt{x+\mu(\rv)/(2T)}\simeq
\sqrt{\mu(\rv)/(2T)}$, the integrand being strongly peaked around
$x=0$). For the same reason, all momenta can be put on the Fermi
level, $p\simeq \sqrt{2m^*(\rv)\mu(\rv)} = p_F = \hbar k_F$. We then
get with $E = \abs{h} = 2T\abs{x}$ and hence $1-2f(E) = \tanh\abs{x}$,
$f'(E) = -1/(4T\cosh^2\abs{x})$, and $f''(E) =
\tanh\abs{x}/(4T^2\cosh^2\abs{x})$ for $\calI_1$ in (\ref{I1})
\begin{equation}
  \calI_1(\rv) =
  \mu(\rv)\frac{N_0(\rv)}{6T^2}\frac{7\zeta(3)}{\pi^2}\,,
  \label{simI1}
\end{equation}
where $ N_0(\rv)=m^*(\rv)k_F(\rv)/(2\pi^2\hbar^2)$ is the local
density of states at the Fermi level per spin. We also used
$\int_0^{\infty}dx \tanh(x)/(x\cosh^2(x)) = 7\zeta(3)/\pi^2$ where
$\zeta(3) \approx 1.202$ is the Riemann zeta function of argument 3.

Let us now compute the other terms of $\kappa(\rv)$ in the limit
$\Delta\ll T$ and under the assumption of weak coupling. For
$\calI_0$, we can use the standard methods from the literature
\cite{Fet71} to get
\begin{equation}
    \calI_0 = N_0(\rv)\bigg[ \ln\frac{8\mu(\rv) e^{\gamma-2}}{\pi T}
      -\frac{7\zeta(3)}{8\pi^2T^2}\abs{\Delta(\rv)}^2\bigg]
      \label{I0GL1}
\end{equation}
(in this equation $\gamma = 0.577$ denotes the Euler constant). Using
the local critical temperature obtained in LDA as $T_c(\rv) =
\Delta_{\text{LDA}}(\rv,T=0)/(\pi e^\gamma)$ [for
  $\Delta_{\text{LDA}}$ see eq. (\ref{gaplda})], eq. (\ref{I0GL1}) can
be cast in the more convenient form
\begin{equation}
    \calI_0 = -\frac{1}{g} + N_0(\rv)\bigg[\ln\frac{T_c(\rv)}{T} -
      \frac{7\zeta(3)}{8\pi^2T^2} \abs{\Delta(\rv)}^2\bigg]\,,
\end{equation}
which agrees with the expression given in \cite{simonucci14} since
$\ln (T_c/T) \approx (T_c-T)/T_c$ for $T$ close to $T_c$. In order to
have a quantitative comparison with the homogeneous infinite matter
situation, let us follow \cite{B-P-98} and see how $\calI_0(\rv)$
varies around the point $\rv=0$. Using $\ln(T_c(\rv)/T) =
\ln(T_c(0)/T)+\ln(T_c(\rv)/T_c(0))$ and $T_c(\rv)/T_c(0) =
\mu(\rv)/\mu(0) e^{1/(gN_0(\rv))-1/(gN_0(0))}$, we get
\begin{multline}
  \calI_0 = -\frac{1}{g}
  + N_0(\rv)\bigg[\ln \frac{T_c(0)}{T} - W(\rv)\\
    -\frac{7\zeta(3)}{8\pi^2}\frac{\abs{\Delta(\rv)}^2}{T^2}\bigg]\,,
  \label{simI_0-2}
\end{multline}
with
\begin{equation}
  W(\rv) = -\bigg[\ln \frac{\mu(\rv)}{\mu(0)}
    + \frac{1}{gN_0(0)}-\frac{1}{gN_0(\rv)}\bigg]\,.
\end{equation}

Inserting now our local pair density, keeping only the $\calI_0$ and
$\calI_1$ terms, into the gap equation (\ref{gapequation}), we obtain
the following GL equation
\begin{multline}
  \bigg[\ln\frac{T_c(0)}{T}-W(\rv)\bigg]\Delta(\rv)
  - \frac{7\zeta(3)}{8\pi^2T^2}\abs{\Delta(\rv)}^2\Delta(\rv)\\
  = -\frac{\mu(\rv)}{6T^2}\frac{7\zeta(3)}{\pi^2}
  \frac{\hbar^2}{4m^*}\nabla^2\Delta(\rv)\,.
  \label{GL-Baranov}
\end{multline}
A very similar equation has been derived earlier by Baranov and Petrov
\cite{B-P-98} in the context of cold atoms where the confining
potential $U(\rv)=m\Omega^2 r^2/2$ (neglecting the Hartree field) is a
harmonic oscillator with trap frequency $\Omega$. In this case, close
to $T_c$, the superfluid phase survives only at small values of $\rv$,
and to lowest order, one obtains the following expression, see
\cite{B-P-98}: $W(r) \simeq r^2/R_{\text{TF}}^2[1+1/(2gN_0(0))]$,
where $R_{\text{TF}} = \sqrt{2\mu(0)/m}/\Omega$ is the Thomas-Fermi
radius. This then leads exactly to eq. (10) of \cite{B-P-98} with the
only difference that there the local chemical potential in the
Laplacian term is replaced by its value at $\rv=0$. Our derivation is
quite different from the one of \cite{B-P-98} and based on a
systematic $\hbar$ expansion of the pairing tensor. It shows that
there exist further gradient terms given below.

Let us now consider the other gradient terms in (\ref{pairdent3}).
We, e.g., want to consider the first term of $c^\kappa_1$. Computing
the corresponding local pair density involves the integral $\int
d^3p/(2\pi \hbar)^3 \tanh(E/(2T))/E^3$. In the limit $T \to T_c$ and,
thus $\Delta \to 0$, this integral diverges. However, taking the sum
of all three terms of $c^\kappa_1$ in (\ref{pairdent3}), making 
again the change of variables $x = h/(2T)$
and proceeding in exactly the same way as in the calculation of
$\calI_1$, we get
\begin{multline}
    \int\!\! \frac{d^3p}{(2\pi\hbar)^3}c^\kappa_1 = -\Delta(\rv)
    \frac{N_0(\rv)}{32 T^2}\\ \times \int_0^\infty dx\bigg[\frac{\tanh
        x}{x^3}-\frac{1}{x^2\cosh^2 x}-\frac{\tanh x}{x\cosh^2
        x}\bigg]\,.
    \label{c1GL}
\end{multline}
Now it is easy to see that the divergences at the Fermi surface
($x=0$) of the first two terms cancel, while the third term has no
divergence at all. Using integration by parts, one can show that the
integral in eq. (\ref{c1GL}) vanishes.

The integral of $c^\kappa_2$ can be neglected because the integrand is
odd in $x$ as mentioned in the calculation of $\calI_1$ above
eq. (\ref{simI1}). The same is true for the integral of $c^\kappa_4$,
whereas the integral of $c^\kappa_5$ gives
\begin{equation}
    \int\!\! \frac{d^3p}{(2\pi\hbar)^3} c^\kappa_5
      = -\frac{N_0}{16T^2} \frac{7\zeta(3)}{\pi^2}\,,
    \label{c5GL}
\end{equation}
and the integral of $c^\kappa_7$ was already discussed in the context
of $\calI_1$.

In conclusion, the only term which additionally enters in the local
pair density is $f_5$. It contributes to the coefficients $Y^\kappa_7$
and $Y^\kappa_9$ [cf. eqs. (\ref{functionsfi}) and
  (\ref{pairden1})]. Replacing under the integral with $c^\kappa_5$
the factor $p^2$ in $f_5$ by $2m^*(\rv)\mu(\rv)$, one finds the
following expressions for these coefficients:
\begin{equation} Y^\kappa_7(\rv) = -K \frac{N_0}{2 T^2}
    \frac{\hbar^2}{m^*}\,,\quad Y^\kappa_9(\rv) = -K
    \frac{N_0\mu(\rv)}{6 T^2}\frac{\hbar^2}{m^*}\,.
\end{equation}
where $K = 7\zeta(3)/(8\pi^2)\simeq 0.1066$. The gap (GL) equation
with these terms added then becomes
\begin{multline}
  \bigg [\ln \frac{T_c(0)}{T} - W(\rv)\bigg ]\Delta
  -K\frac{\abs{\Delta}^2}{T^2}\Delta \\
  - \frac{K}{2T^2}\frac{\hbar^2}{m^*}
    \nabla U\cdot \nabla \Delta
  - \frac{K\mu(\rv)}{6T^2} \hbar^2 \nabla \Big(\frac{1}{m^*}\Big)
    \cdot \nabla \Delta\\
    = -\frac{4K\mu(\rv)}{3T^2} \frac{\hbar^2}{4m^*}
      \nabla^2\Delta
  \label{GLE-f}
\end{multline}
This completes the derivation of the most general GL equation for
finite Fermi systems with a local mean field and effective mass
generated from a complete expansion of the pairing tensor at finite
temperature to order $\hbar^2$.

Notice that in the case of a harmonic trap and no effective mass,
considered in \cite{B-P-98}, the $f_5$ term becomes $f_5 = \nabla
U\cdot\nabla \Delta = m\Omega^2 r\, d\Delta(r)/dr$.

\section{Application to atoms in a harmonic trap}
\label{sec:results} 
Let us now apply the semiclassical equations derived in the previous
sections to the case of a Fermi gas with attractive interaction in a
harmonic trap. In experiments with cold atoms, the trap has usually a
cylindrical shape, with $\Omega_z\ll \Omega_x,\Omega_y$. In this
situation, non-negligible corrections to the LDA might come from the
strong gradients in $x$ and $y$ directions. Nevertheless, in the
present work, we will restrict ourselves to the spherical case
$\Omega_x = \Omega_y = \Omega_z = \Omega$, because we wish to compare
our semiclassical results to the solution of the fully quantum
mechanical HFB equations, which are only available in spherical
symmetry.

We solve the HFB equations as in \cite{grasso03}, but without the
Hartree field. Notice that the simple Hartree field of the form
$U_{\text{Hartree}}(\rv) = g\rho(\rv)/2$ is only valid at weak
coupling. At stronger coupling, it must be replaced by the real part
of the self-energy, computed, e.g., with the in-medium T matrix in
ladder approximation, which effectively weakens the interaction
\cite{Chiacchiera09}. To avoid such complications, we will neglect the
Hartree field and keep only the external trap potential $U(\rv) =
m\Omega^2 r^2/2$. We will furthermore set $m^* = m$ (i.e., $\gamma =
1$).

Let us briefly outline how we solve the nonlinear differential
equation (\ref{hfb1}) and its generalization when all $Y^\kappa_i$
coefficients are included in eq. (\ref{pairden1}) for $\kappa_2$. In
spherical symmetry, for a given potential $U(r)$ and without effective
mass, this equation can be written in the form
\begin{equation}
  Y^\kappa_0 + Y^\kappa_1 \nabla^2\Delta + Y^\kappa_7 U'\Delta' = 0\,,
  \label{nonlin}
\end{equation}
where $\Delta'= d\Delta/dr$, $U' = dU/dr$, $\nabla^2 \Delta = (r
\Delta)''/r$, and the $Y^\kappa_i$ are only functions of $r$,
$\Delta$, and $T$. The coefficient $Y^\kappa_0$ combines all terms
that do not involve any derivatives of $\Delta$, i.e.,
\begin{equation}
  Y^\kappa_0 = \frac{\Delta}{g} + \calI_0 + Y^\kappa_3 \nabla^2 U +
  Y^\kappa_4 U'^2\,.
\end{equation}

Analytical expressions for the coefficients $Y^\kappa_i$ at $T=0$ are
given in \cite{simonucci14} and in the appendix
\ref{app:yi}. Analytical expressions exist also in the GL limit
$\Delta \ll T \ll \mu(r)$ and are given in \cite{simonucci14} and in
sect. \ref{sec:gl}. In the general $T>0$ case, however, we have to
perform the momentum integrals of the functions given in
eq. (\ref{pairdent2}) numerically, carefully sampling in particular
the regions $\abs{h}\lesssim T$ and $\abs{h}\lesssim \Delta$.

By discretizing $\Delta(r)$ on a radial mesh, we transform
eq. (\ref{nonlin}) into a system of coupled equations. In our
calculations, we use 4- and 5-point rules for the first derivative and
Laplacian, respectively, with special rules at the end points $r=0$
and $r=r_{\text{max}}$. This system of equations is nonlinear
because the coefficients $Y^\kappa_i$ depend on $\Delta$, and is
solved iteratively, with a method similar to a damped Newton method.

In the presentation and discussion of the results, it is convenient to
consider so-called `trap units' in which $\hbar = \Omega = m = 1$. In
practice, this means that energies are measured in units of
$\hbar\Omega$, lengths in units of $\sqrt{\hbar/(m\Omega)}$, and so
on. These units will be used throughout this section.

\begin{figure}
  \includegraphics[width=7.5cm]{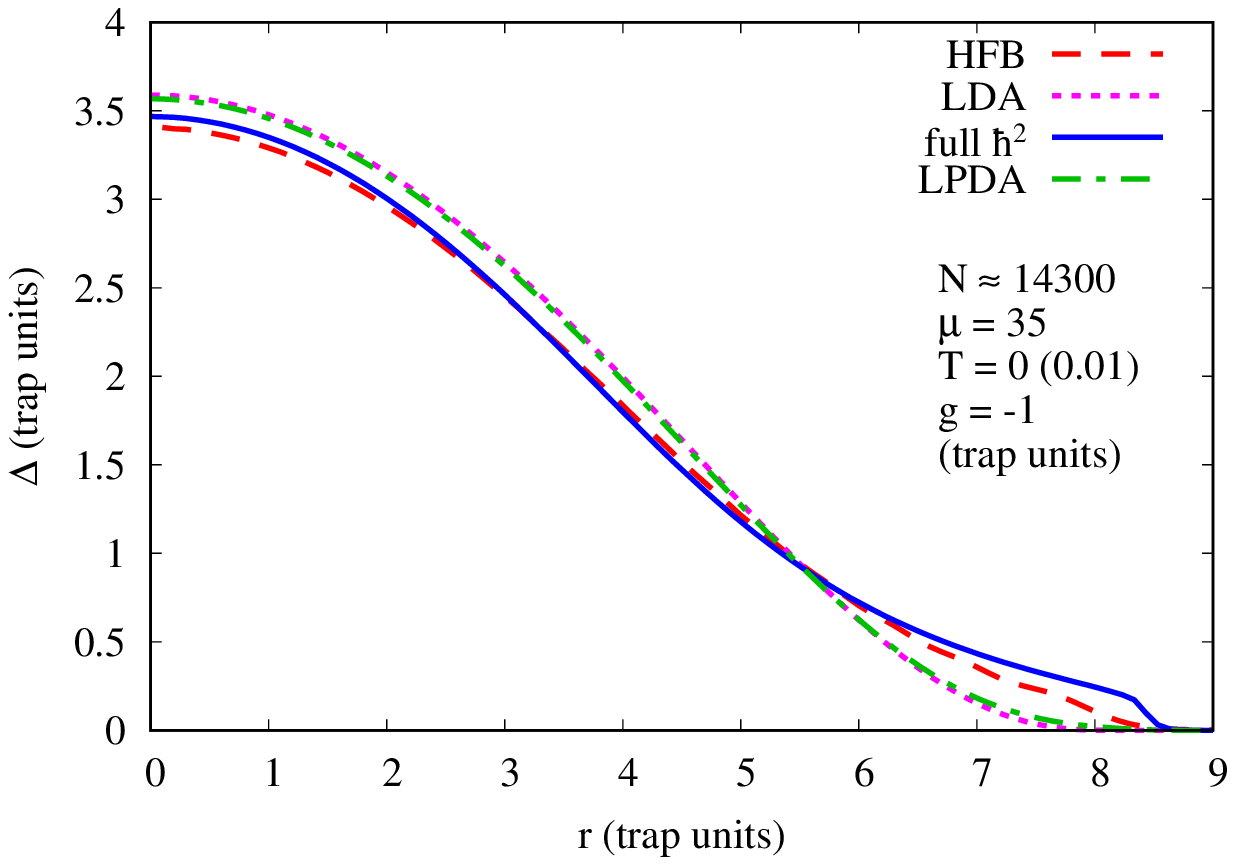}\\
  \includegraphics[width=7.5cm]{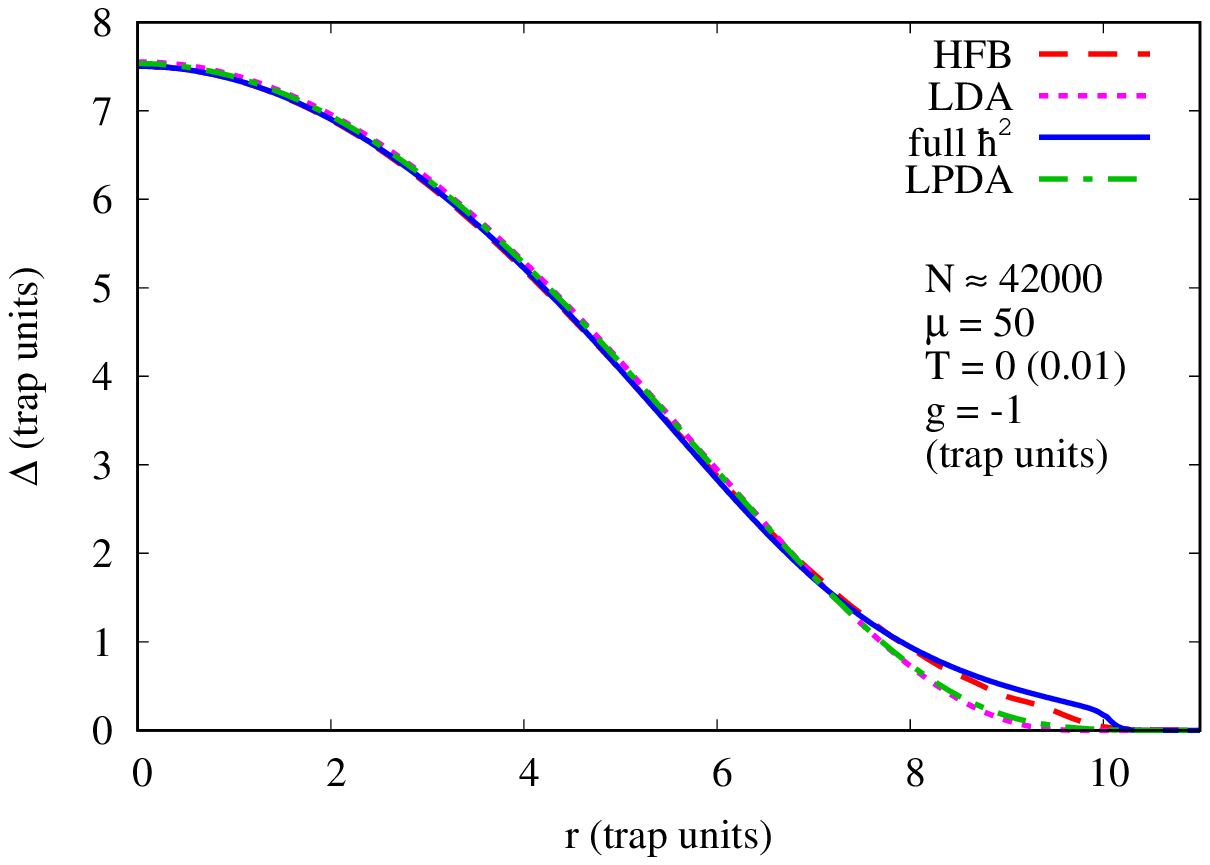}
  \caption{Results for the gap $\Delta$ obtained at different levels
    of approximation as a function of $r$ at $T=0$ for $\mu = 35$
    (upper panel) and $\mu = 50$ (lower panel) in trap units (i.e.,
    $r$ in units of $\sqrt{\hbar/(m\Omega)}$, $\mu$, $\Delta$, and $T$ in
    units of $\hbar\Omega$).
    \label{fig:T0}}
\end{figure}

Let us start with $T=0$. Figure \ref{fig:T0}
displays the $r$ dependence of the gap computed in different approximations, namely
HFB (red long dashed lines), LDA (purple dotted lines), the full $\hbar^2$ 
correction (blue solid lines), and LPDA (green dash-dot lines), for two cases: 
$\mu = 35$ (top) and $\mu = 50$ (bottom), corresponding approximately to particle
numbers $14300$ and $42000$, respectively (the precise number depends on
which approximation is employed for the gap). The coupling constant is
fixed to $g=-1$, and hence the often used dimensionless interaction
strength parameter $k_F(r=0)a_F\approx \sqrt{2\mu}g/(4\pi)$ is $-0.67$
in the upper panel and $-0.80$ in the lower one. Let us first look at
the HFB results and compare them with the results obtained at leading order in
$\hbar$, i.e., the LDA. In the larger system ($\mu=50$), the LDA works very
well except near the surface, while in the smaller system ($\mu=35$)
we see that the LDA overestimates the gap in the center. Near the
surface, in both cases, the LDA gap goes to zero too rapidly as $r$
approaches the classical turning point $R_{\text{TF}}$. In HFB, the
gap actually extends slightly beyond $R_{\text{TF}}$. Similar
observations were already made in \cite{grasso03}.

Let us now see how the LDA result is improved by the LPDA and by the full $\hbar$
expansion to order $\hbar^2$. We see that on the scale of the graphs, the LPDA
results are hardly distinguishable from the LDA ones, while the full
$\hbar^2$ corrections bring a clear improvement compared to the LDA:
In the $\mu = 35$ case, we see that the reduction of the gap in the
center is fairly well reproduced by the $\hbar^2$ calculation. Also
near the surface, the $\hbar^2$ corrected gap follows more closely the
HFB gap than the LDA, it even becomes too large around the classical
turning point. We should point out, however, that because of the
divergence of some of the $Y^\kappa_i$ coefficients at $T=0$ in the
limit $\Delta\to 0$, we cannot perform the $\hbar^2$ calculation
exactly at $T=0$ but we have to do it at a small but finite
temperature ($T=0.01$). The results do not show a pronounced
dependence on the chosen value of this small temperature, e.g., with
$T=0.1$ we obtain almost the same curves as with $T=0.01$.

\begin{figure}
  \includegraphics[width=7.5cm]{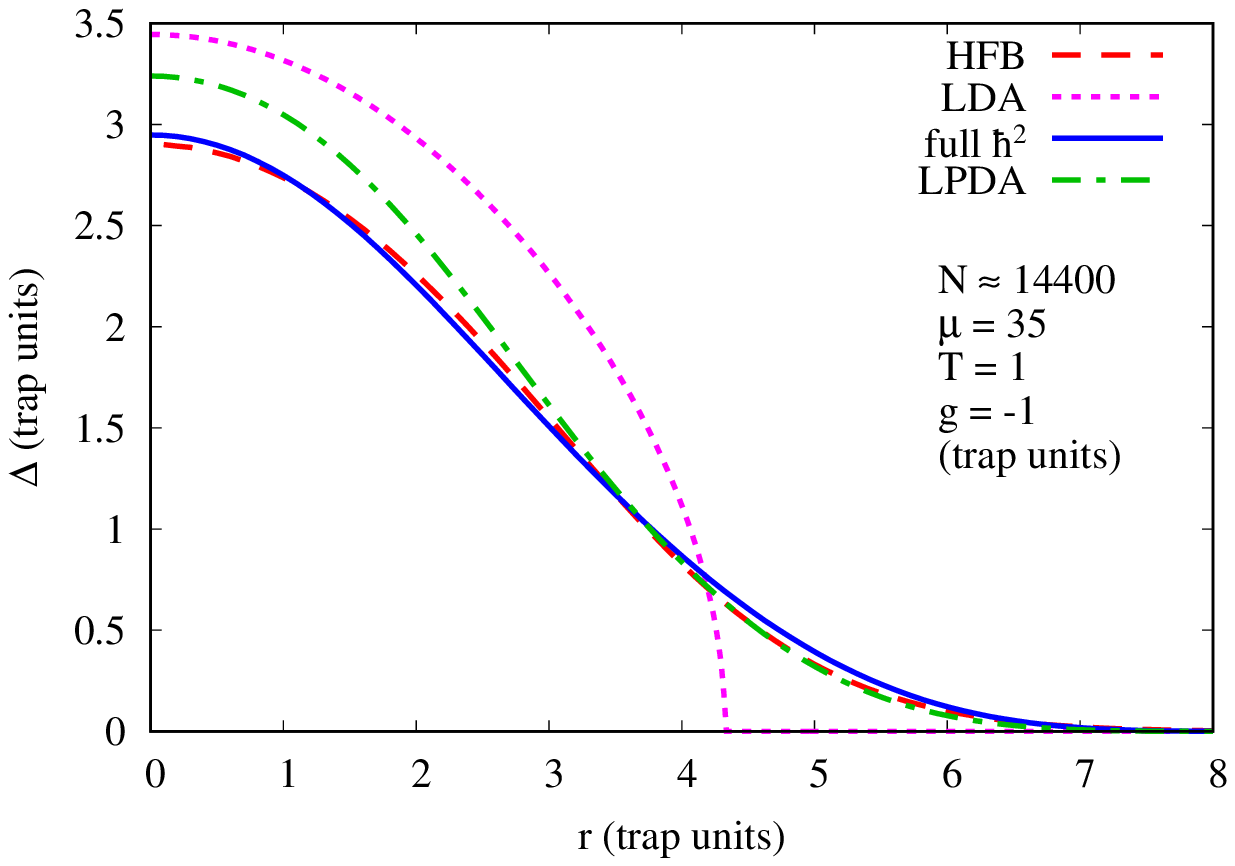}\\
  \includegraphics[width=7.5cm]{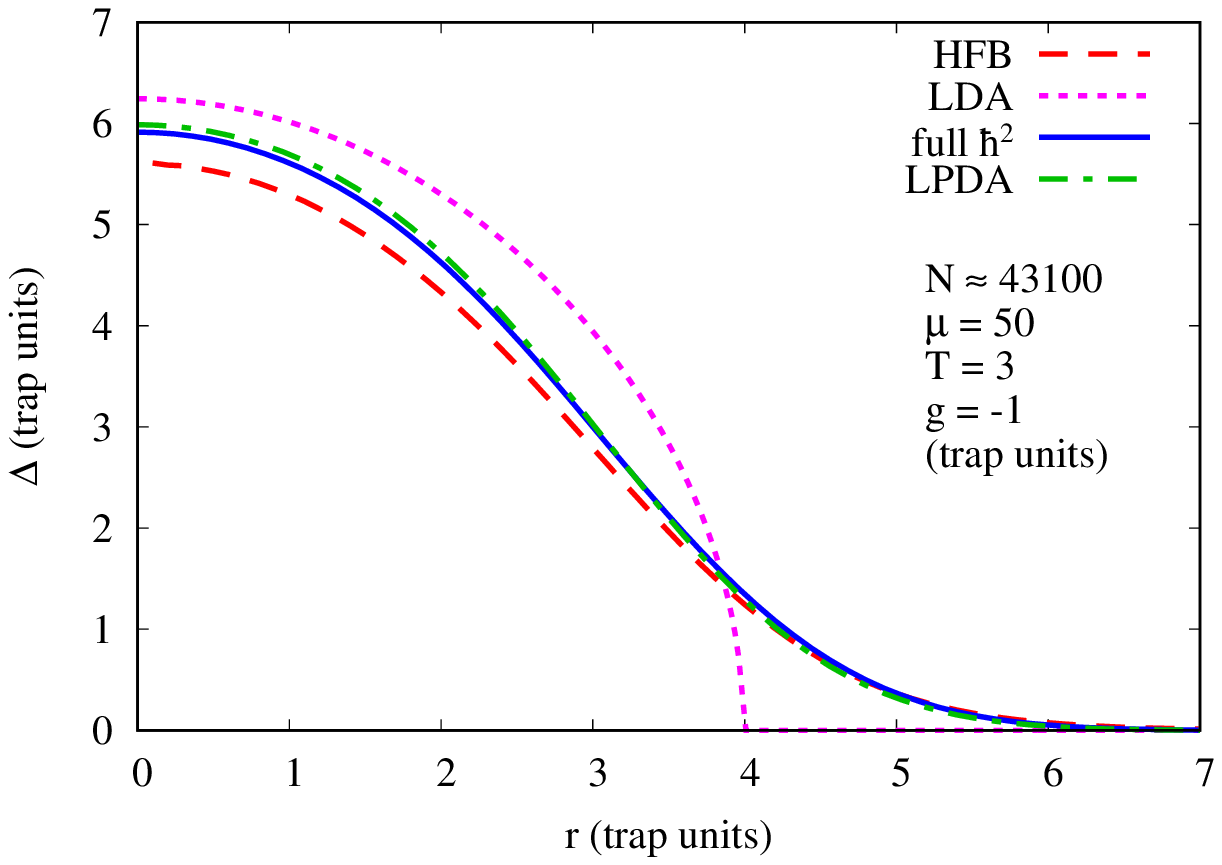}
  \caption{Same as fig. \ref{fig:T0} but at temperature $T=1$ in the
    case $\mu = 35$ (upper panel) and $T=3$ in the case $\mu =50$
    (lower panel).
    \label{fig:T1}}
\end{figure}

In fig. \ref{fig:T1},
we consider again the same systems, but now at finite temperature:
$T=1$ in the case $\mu = 35$ (top) and $T=3$ in the case $\mu = 50$
(bottom). As already observed in \cite{grasso03}, the LDA fails badly
at finite temperature. Compared to the HFB result, the LDA gap is too
large in the center, and it drops too abruptly to zero at the radius
where $T_c(r) = T$, while the HFB gap goes smoothly to zero at a much
larger radius. We see that both the LPDA and the full $\hbar^2$
calculations are quite successful in reproducing the general behavior
of the HFB gap, both in the center and in the tail.

\begin{figure}
  \includegraphics[width=7.5cm]{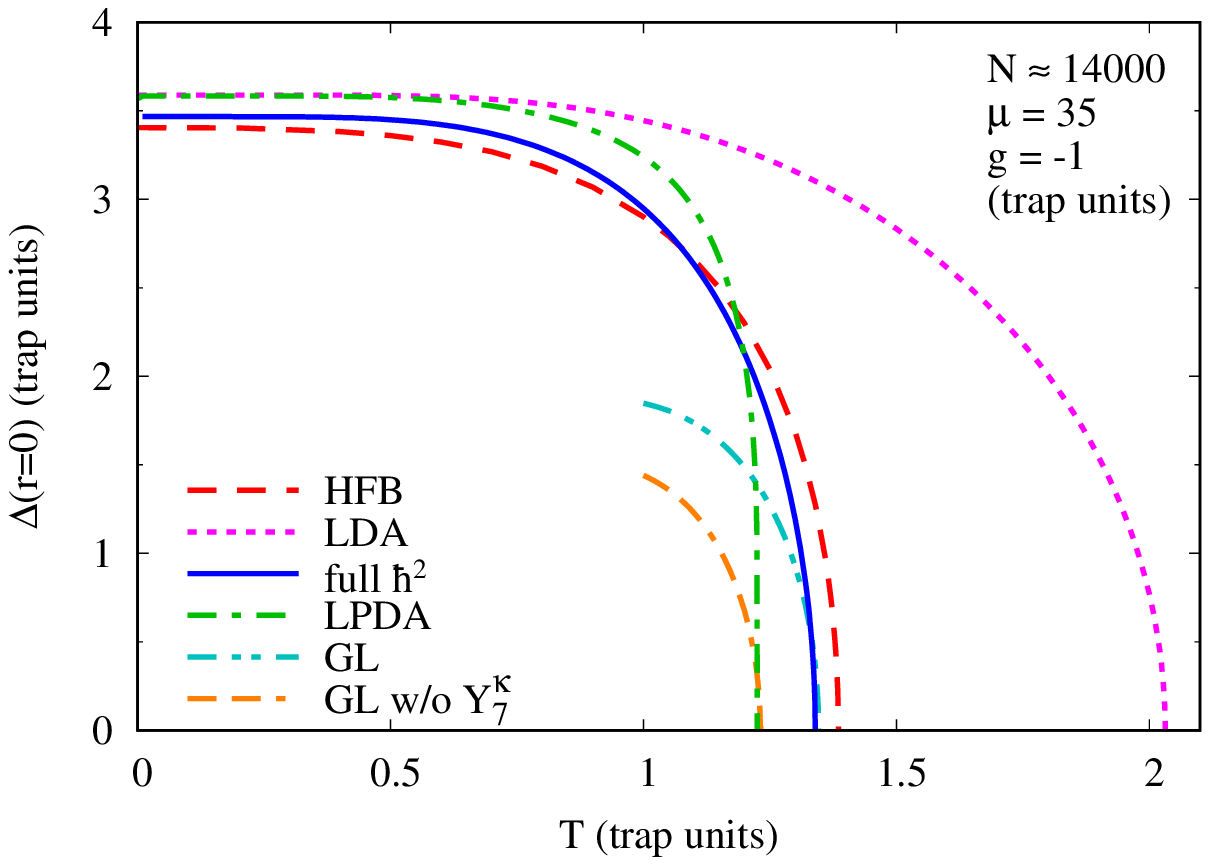}\\
  \includegraphics[width=7.5cm]{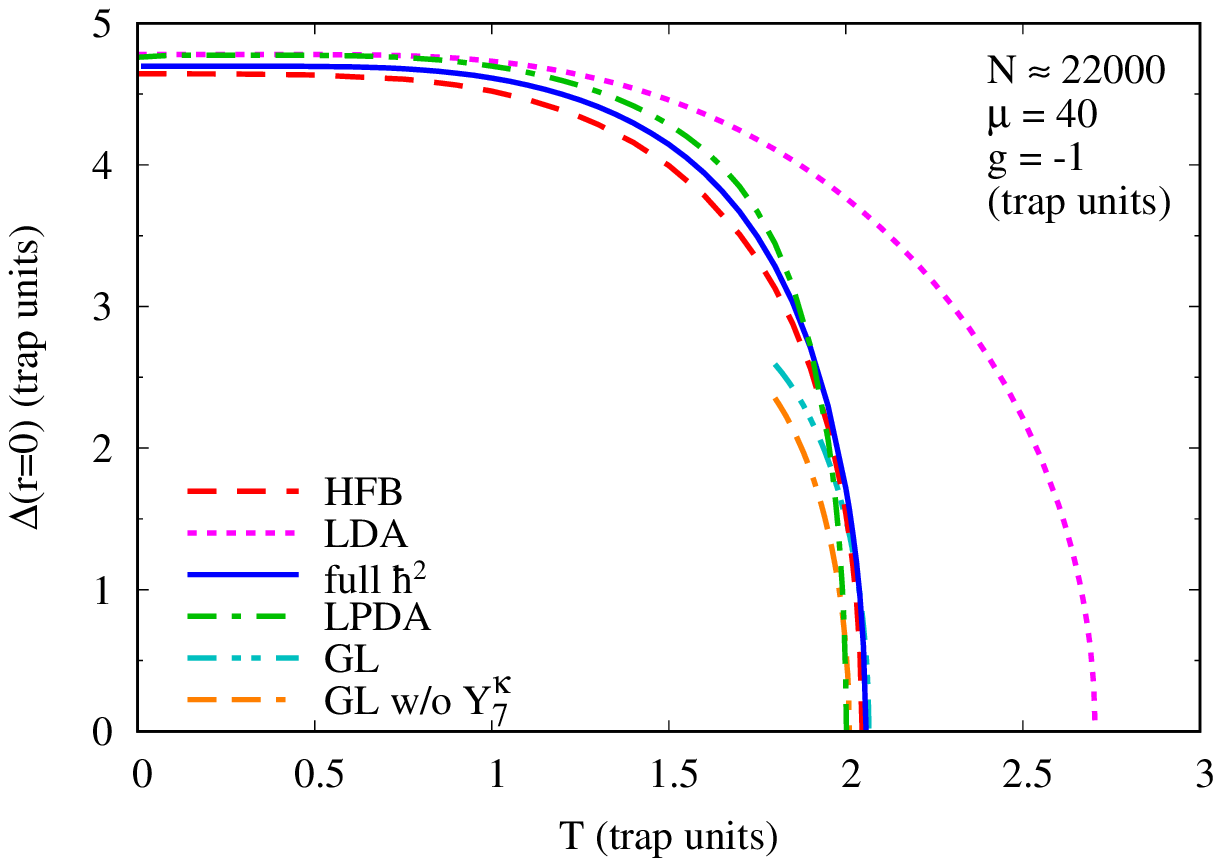}\\
  \includegraphics[width=7.5cm]{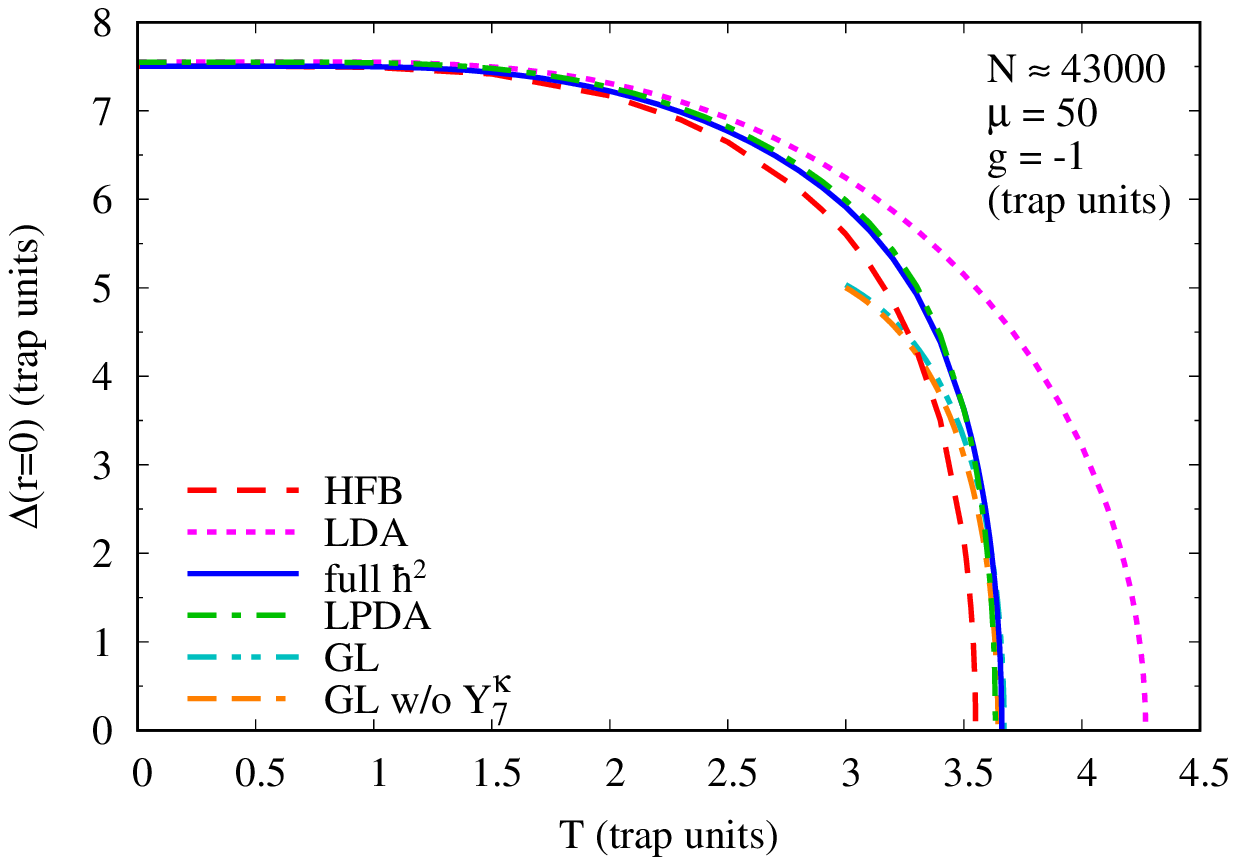}
  \caption{Value of the gap in the center $(r=0)$ as a function of
    temperature for three different systems (from top to bottom: $\mu
    = 35$, $40$, and $50$) at different levels of approximation.
  \label{fig:Delta0}}
\end{figure}

The fact that in the upper panel of fig. (\ref{fig:T1}) the full
$\hbar^2$ calculation agrees better with the HFB result than the LPDA,
while in the lower panel the $\hbar^2$ results are very close to the
LPDA one, should not be overinterpreted as this depends sensitively on
the chosen temperature and chemical potential. This can be seen in
fig. \ref{fig:Delta0},
where we display the value of the gap at the center, $\Delta(r=0)$, as
a function of the temperature, for three different chemical
potentials: $\mu = 35$ (top), $40$ (middle), and $50$
(bottom). Comparing with the HFB calculation as a reference (red
dashed lines), we see again the failure of the LDA (purple dotted
lines) in all three cases, predicting not only a larger gap than HFB,
but also a critical temperature $T_c$ that is clearly too high. Both
the full $\hbar^2$ calculation (solid blue line) and LPDA (green
dash-dotted line) are able to bring $T_c$ down to values that are
close to the $T_c$ obtained in HFB. Notice, however, that especially
at $\mu=35$, the rise of the gap $\Delta(r=0)$ when the temperature is
lowered below $T_c$ is too steep, and also, unlike the full $\hbar^2$
calculation, the LPDA always has a gap very close to the LDA one at
$T=0$.

To get the value of $T_c$, it is actually not necessary to use the
numerical coefficients $Y^\kappa_i$, but one can also use the
analytical coefficients in the corresponding GL limit (turquoise
dash-dot-dot lines: GL equation corresponding to the full
$\hbar^2$ expansion including $Y^\kappa_1$ and $Y^\kappa_7$, and
orange dash-dash-dot line: GL equation corresponding to the LPDA,
i.e., without $Y^\kappa_7$, as in \cite{B-P-98,simonucci14}).

\begin{figure}
  \includegraphics[width=7.5cm]{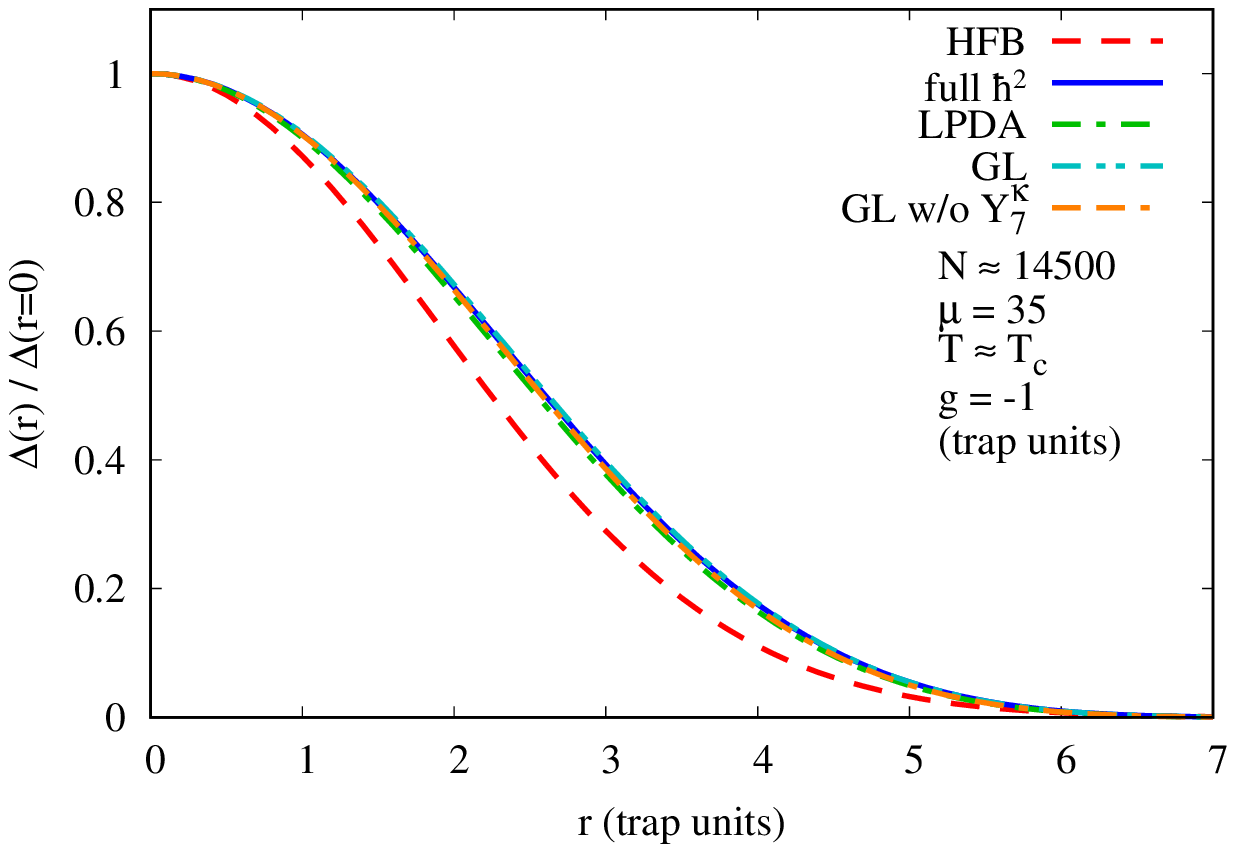}\\
  \includegraphics[width=7.5cm]{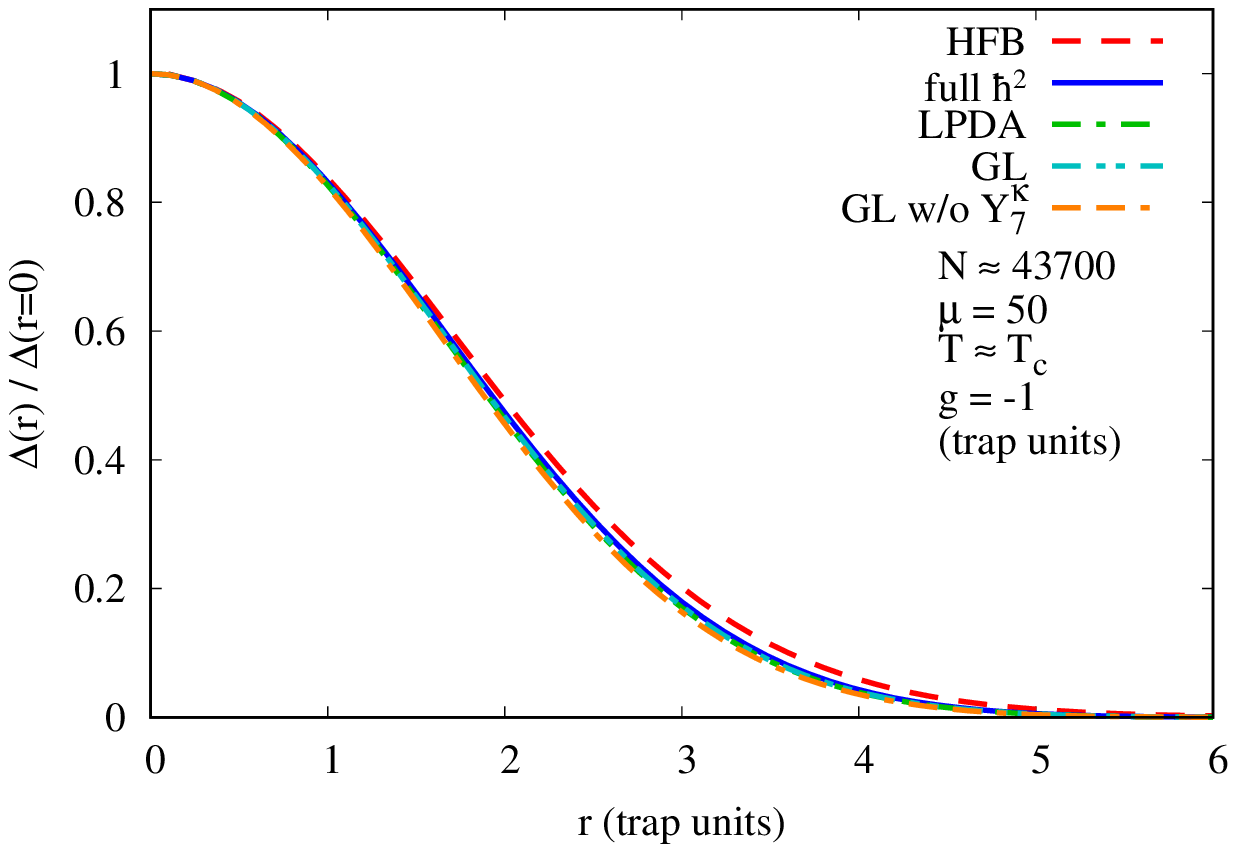}
  \caption{Results for $\Delta(r)/\Delta(r=0)$ at different levels of
    approximation, in the limit that in each approximation the
    temperature approaches the respective critical temperature $T_c$,
    for $\mu = 35$ (upper panel) and $\mu = 50$ (lower panel).
    \label{fig:Tc}}  
\end{figure}

The reduction of $T_c$ as compared to LDA was for the first time
discussed in \cite{B-P-98} in the context of the GL equation
(\ref{GL-Baranov}) including only the Laplacian term,
$Y^\kappa_1$. There, it was also discussed that close to $T_c$, the
gap $\Delta(r)$ has the shape of a Gaussian whose width remains finite
and only the magnitude goes to zero when $T\to T_c$. To connect with
this study, we display in fig. \ref{fig:Tc}
the functions $\Delta(r)/\Delta(r=0)$ for each approximation [full
  $\hbar^2$ (blue solid line, LPDA (green dash-dotted line, GL with
  $Y^\kappa_7$ (turquoise dash-dot-dot line), and GL without
  $Y^\kappa_7$ (orange dash-dash-dot line)] in the limit that $T$
approaches the respective $T_c$, in the two cases $\mu = 35$ (top) and
$\mu = 50$ (bottom). Somewhat surprisingly, we see that in both cases,
the shapes of $\Delta(r)$ in the different approximations agree very
well among each other but not so well with the HFB (red dashed
lines). In particular, the disagreement gets worse in the smallest
system, $\mu=35$, indicating that the $\hbar$ expansion might approach
its limit of applicability here. This could have been guessed from the
very large correction to $T_c$ compared to the LDA one in this
case. In fact, one expects that in very small systems (or systems with
very weak pairing), the $\hbar$ expansion should fail when the
coherence length of the Cooper pairs becomes comparable with the
system size.
\section{\label{sec:conclusions} Conclusions and Discussion}
In this work, we took up the semiclassical theory for the radius
dependence of the local gap function of finite and inhomogeneous Fermi
systems which was obtained from a coarse graining of the quantal
equations in \cite{simonucci14}. The approach which lead to a second
order differential equation for the local gap, named LPDA, showed for
many situations of cold atom systems very good agreement with full
solutions of the HFB or BdG equations \cite{simonucci14}. The aim of this paper was 
to compare the LPDA with the expression of the anomalous density matrix 
obtained long ago in a systematic $\hbar$ expansion up to order $\hbar^2$
\cite{taruishi92}. It was revealed that one of the many
terms of the full $\hbar$ expansion resembles
the LPDA. This term, proportional to
$\nabla^2\Delta$, is actually the most important one, as was confirmed by our numerical studies. However, the coefficients of the $\nabla^2\Delta$ term in the $\hbar$ expansion and in the LPDA are different.

We also considered the GL regime. Close to the critical temperature,
our full expression for the anomalous density reduces to an
analytical expression. We compared our GL equation with the one obtained by
Baranov and Petrov \cite{B-P-98}, which was derived for cold atoms
confined in harmonic potentials, and found that our equation contains
an additional term. In the GL limit, the coefficients of the $\nabla^2\Delta$
terms in the $\hbar$ expansion and in the LPDA agree with each other.

There remain some open questions that we could not address in the present paper. In particular, we considered the $\hbar$ expansion only for the case of a real gap. However, to describe dynamical processes, the phase of the complex gap is crucial. In the $\hbar$ expansion of the time-dependent HFB equation, the phase must be treated separately to avoid the mixing of even and odd orders in $\hbar$ \cite{Urb06}. In the special case of a stationary rotation, a generalization of the LDA was employed in ref. \cite{Urb08} that treated the gradient of the phase (i.e., the superfluid velocity) exactly and neglected only gradients of the modulus of $\Delta$. Similarly, the LPDA was specifically derived to describe vortices in a rapidly rotating Fermi gas by expanding the solution of the BdG equation for a spatially constant superfluid velocity. The different coefficients of the $\nabla^2 \Delta$ term in the $\hbar$ expansion and in the LPDA might come from the way how the gradients of the phase survive in the final equation when one considers the case of a real gap. This statement is, however, still speculative and requires more careful investigations.
\section*{Acknowledgments}
We are very greatful to G. C. Strinati who introduced us to many
details of the derivation of the formulas in \cite{simonucci14}. One
of us, X.V., was partially supported by Grants No.\ FIS2017-87534-P
from MINECO and No.\ CEX2019-000918-M from AEI-MICINN through the
``Unit of Excellence Mar\'{\i}a de Maeztu 2020-2023'' award to ICCUB.
\appendix
\section{Explicit expressions for the $Y^\kappa_i$ and $Y^\rho_i$
  coefficients at $T=0$}
\label{app:yi}
As we have mentioned in the main text, the $\pv$ integrals of the pair
and normal density matrices (\ref{pairdent2}) and (\ref{normdent2})
can be performed analytically at zero temperature. However, it will
turn out that there are certain terms which are divergent in the limit
$\Delta \to 0$. We report here the corresponding values $Y^\kappa_i$
as they are obtained from Pei (including all divergent terms)
\cite{pei15}.

The functions $Y^\kappa_i$ and $Y^\rho_i$ can be expressed in terms of
two functions $I_5(x_0)$ and $I_6(x_0)$, where $x_0=\mu(\rv)/\Delta(\rv)$,
which in turn can be written in terms of two complete elliptic
integrals as follows \cite{marini98,simonucci14}:
\begin{gather}
I_5(x_0) = (1 + x_0^2)^{1/4} E\big(\tfrac{\pi}{2},\kappa\big)
-\frac{F\big(\tfrac{\pi}{2},\kappa\big)}{4 x_1^2(1+x_0^2)^{1/4}}
\label{I5}\\
I_6(x_0) = \frac{1}{2(1+x_0^2)^{1/4}} F\big(\tfrac{\pi}{2},\kappa\big),
\label{I6}
\end{gather}
where $x_1^2 = (\sqrt{1 + x_0^2} + x_0)/2$ and
$\kappa^2=x_1^2/\sqrt{1+x_0^2}$, while $F(\pi/2,\kappa)$ and
$E(\pi/2,\kappa)$ are the complete elliptic integrals of first and
second kind defined by \cite{abramowitz72,gradshteyn65}
\begin{gather}
  F(\alpha,\kappa) = \int_0^{\alpha} d\phi
    \frac{1}{\sqrt{1 - \kappa^2 \sin^2 \phi}}\,,\label{F}\\
  E(\alpha,\kappa) = \int_0^{\alpha} d\phi \sqrt{1 - \kappa^2 \sin^2 \phi},
\label{E}
\end{gather}
with $\kappa^2 < 1$. The main properties of these elliptic integrals
are given in appendix A of \cite{marini98}.

If we write for brevity $I_5$ and $I_6$ for $I_5(x_0)$ and $I_6(x_0)$,
respectively, the analytical expressions for the $Y^\kappa_i$
functions read
\begin{multline}
  Y^\kappa_1(\rv) =
  \frac{1}{144\pi^2}\Big(\frac{2m^*}{\hbar^2}\Big)^{1/2}
    \frac{1}{\sqrt{\Delta}}\\
  \times \frac{(10x_0^2+7)I_6+(4x_0^2+1)x_0I_5}{1+x_0^2}\,,
\label{Y1}
\end{multline}
\begin{multline}
  Y^\kappa_2(\rv) = -\frac{1}{576\pi^2}\Big(\frac{2m^*}{\hbar^2}\Big)^{1/2}
  \frac{1}{\Delta\sqrt{\Delta}}\\
  \times \frac{(4x_0^4+23x_0^2+7)I_6+(16x_0^4+38x_0^2+10)x_0I_5}{(1+x_0^2)^2}\,,
\label{Y2}
\end{multline}
\begin{multline}
Y^\kappa_3(\rv) = -\frac{1}{48\pi^2}\Big(\frac{2m^*}{\hbar^2}\Big)^{1/2}
\frac{1}{\sqrt{\Delta}}\frac{I_5-x_0I_6}{1+x_0^2}\,,
\label{Y3}
\end{multline}
\begin{multline}
Y^\kappa_4(\rv) = \frac{1}{192\pi^2}\Big(\frac{2m^*}{\hbar^2}\Big)^{1/2}
\frac{1}{\Delta\sqrt{\Delta}}\\
\times \frac{(3x_0^2-1)I_6-4x_0I_5}{(1+x_0^2)^2}\,,
\label{Y4}
\end{multline}
\begin{equation}
Y^\kappa_5(\rv) = - \frac{1}{24\pi^2}\Big(\frac{2m^*}{\hbar^2}\Big)^{1/2}
\sqrt{\Delta}I_6\,,
\label{Y5}
\end{equation}
\begin{equation}
Y^\kappa_6(\rv) =  \frac{7}{192\pi^2}\Big(\frac{2m^*}{\hbar^2}\Big)^{1/2}
\sqrt{\Delta}I_6\,,
\label{Y6}
\end{equation}
\begin{multline}
  Y^\kappa_7(\rv) = - \frac{1}{96\pi^2}\Big(\frac{2m^*}{\hbar^2}\Big)^{1/2}
  \frac{1}{\Delta\sqrt{\Delta}}\\
  \times \frac{(4x_0^4+11x_0^2+3)I_5-(2x_0^2-2)x_0I_6}{(1+x_0^2)^2}\,,
\label{Y7}
\end{multline}
\begin{equation}
Y^\kappa_8(\rv) = \frac{1}{96\pi^2}\Big(\frac{2m^*}{\hbar^2}\Big)^{1/2}
\frac{1}{\sqrt{\Delta}}\frac{I_5-x_0I_6}{1+x_0^2}\,,
\label{Y8}
\end{equation}
\begin{multline}
  Y^\kappa_9(\rv) = - \frac{1}{96\pi^2}\Big(\frac{2m^*}{\hbar^2}\Big)^{1/2}
  \frac{1}{\sqrt{\Delta}} \\
  \times \frac{(10x_0^2+7)I_6-(4x_0^2+1)x_0I_5}{1+x_0^2}\,,
\label{Y9}
\end{multline}

In our study of the pairing in finite systems it is actually relevant
to know the $\Delta \to 0$ limit because the gap goes to zero at the
surface. In this case the auxiliary functions $I_5(x_0)$ and
$I_6(x_0)$ behave as $I_5(x_0) \simeq \sqrt{x_0} \simeq
1/\sqrt{\Delta}$ and $I_6(x_0)\simeq \ln (8x_0)/(2 \sqrt{x_0}) \simeq
\sqrt{\Delta}\ln \Delta$, respectively \cite{marini98}. This implies
that, as already mentioned, in the limit $\Delta \to 0$ some of the
$Y^\kappa_i$ functions defined previously are divergent. For example,
in the function $Y^\kappa_1$ (\ref{Y1}) the leading term in this limit
behaves as $\nabla^2 \Delta/\Delta^2$ and therefore diverges.

It is easy to show that the $\hbar^0$ (Thomas-Fermi) contribution to
the normal density in the presence of the pairing field is given
by \cite{marini98,simonucci14}
\begin{multline}
\rho_0(\rv) = \int\!\! \frac{d^3p}{(2\pi\hbar)^3}
\frac{1}{2}\bigg[ 1 - \frac{h(\rv,\pv)}{E(\rv,\pv)}\bigg] \\
  = \frac{1}{6\pi^2} \bigg(\frac{2m^*\Delta}{\hbar^2}\bigg)^{3/2}[I_6
    + x_0 I_5]\,.
\end{multline}

The contributions to the $\hbar^2$ corrections of the normal density
$\rho_2$ as given in Pei \cite{pei15} read
\begin{equation}
Y^\rho_1(\rv) =  \frac{1}{48\pi^2}\Big(\frac{2m^*}{\hbar^2}\Big)^{1/2}
\frac{1}{\sqrt{\Delta}}\frac{(x_0^2+3)I_5-3x_0I_6}{1+x_0^2}\,,
\label{Z1}
\end{equation}
\begin{multline}
Y^\rho_2(\rv) = - \frac{1}{192\pi^2}\Big(\frac{2m^*}{\hbar^2}\Big)^{1/2}\frac{1}{\Delta\sqrt{\Delta}}\\
\times \frac{(4x_0^4+5x_0^2+5)I_5+(2x_0^2-2)x_0I_6}{(1+x_0^2)^2}\,,
\label{Z2}
\end{multline}
\begin{equation}
Y^\rho_3(\rv) = - \frac{1}{48\pi^2}\Big(\frac{2m^*}{\hbar^2}\Big)^{1/2}
\frac{1}{\sqrt{\Delta}}\frac{I_6+x_0I_5}{1+x_0^2}\,,
\label{Z3}
\end{equation}
\begin{multline}
Y^\rho_4(\rv) = - \frac{1}{192\pi^2}\Big(\frac{2m^*}{\hbar^2}\Big)^{1/2}
  \frac{1}{\Delta\sqrt{\Delta}}\\
  \frac{(x_0^2-3)I_5+4x_0I_6}{(1+x_0^2)^2}\,,
\label{Z4}
\end{multline}
\begin{equation}
Y^\rho_5(\rv) = - \frac{1}{24\pi^2}\Big(\frac{2m^*}{\hbar^2}\Big)^{1/2}
\sqrt{\Delta}I_5\,,
\label{Z5}
\end{equation}
\begin{equation}
Y^\rho_6(\rv) = \frac{7}{192\pi^2}\Big(\frac{2m^*}{\hbar^2}\Big)^{1/2}
\sqrt{\Delta}I_5\,,
\label{Z6}
\end{equation}
\begin{multline}
Y^\rho_7(\rv) = - \frac{1}{96\pi^2}\Big(\frac{2m^*}{\hbar^2}\Big)^{1/2}
\frac{1}{\Delta\sqrt{\Delta}}\\
\frac{(3x_0^2-1)I_6-4x_0I_5}{(1+x_0^2)^2}\,,
\label{Z7}
\end{multline}
\begin{equation}
Y^\rho_8(\rv) = \frac{1}{96\pi^2}\Big(\frac{2m^*}{\hbar^2}\Big)^{1/2}
\frac{1}{\sqrt{\Delta}}\frac{I_6+x_0I_5}{1+x_0^2}\,,
\label{Z8}
\end{equation}
\begin{equation}
Y^\rho_9(\rv) = - \frac{1}{96\pi^2}\Big(\frac{2m^*}{\hbar^2}\Big)^{1/2}
\frac{1}{\sqrt{\Delta}}\frac{I_5-3x_0I_6}{1+x_0^2}\,.
\label{Z9}
\end{equation}

In the $\Delta \to 0$ limit, only the $Y^\rho_3$, $Y^\rho_4$,
$Y^\rho_5$, $Y^\rho_6$, and $Y^\rho_8$ terms in $\rho_2(\rv)$ survive,
because the others are multiplied by gradients of $\Delta$. If we
furthermore consider $m^* = m$, only $Y^\rho_3$ and $Y^\rho_4$ are
relevant. Taking into account the asymptotic behaviour of $I_5(x_0)$
and $I_6(x_0)$ in the limit $\Delta\to 0$, the normal density in this
case agrees with the well-known normal density given by eq. (13.44) of
ref. \cite{Rin80}.
\section{$2\times 2$ generalized density matrix from
  ref. \cite{taruishi92}}
\label{app:R}
The finite temperature expressions of section \ref{sec:hbar2}
can be straightforwardly derived from eqs. (5.5) and (5.7) of
\cite{taruishi92}. The zeroth order of the $2\times2$ generalized
  density matrix is given by
\begin{equation}
  \calR_0 = \frac{1}{2}\bigg [ \Id + \frac{\calH}{E}(1-2f(E)) \bigg ]
\end{equation}
Proceeding with the expression (5.7) in \cite{taruishi92} in the same way,
we obtain
\begin{multline}
  \calR_2(E) =
  \bigg [ g_1\frac{\calH}{E} - g_2\frac{\calF}{E}\bigg ](1-2f(E))\\
  +\bigg [ g_7\frac{\calH}{E} + g_8\frac{\calF}{E}\bigg]
    2\frac{\partial^2f(E)}{\partial E^2}
  + g_{10}\frac{\calH}{E}
    2\frac{\partial^3 f(E)}{\partial E^3}\,,
\end{multline}
where
\begin{equation}
  \calH = \begin{pmatrix} h&\Delta \\ \Delta &-h\end{pmatrix}
     \,,\qquad
  \calF = \begin{pmatrix} -\Delta&h \\ h&\Delta\end{pmatrix}\,.
\end{equation}
Finally, this leads to eqs. (\ref{pairdent2}) and (\ref{normdent2}) of
the main text with the $g_i$ expressed in terms of the $f_i$ given in
\cite{taruishi92}. Please notice that in \cite{taruishi92} there is
a sign misprint and $g_2$ should be replaced by $-g_2$ there.

\end{document}